\documentclass[aps,prl,superscriptaddress,notitlepage,amsmath,amssymb,nofootinbib,longbibliography]{revtex4-1}
\usepackage[utf8]{inputenc} 
\usepackage{graphicx}
\usepackage{url}
\usepackage[bookmarks, pagebackref=false]{hyperref}
\usepackage[usenames,dvipsnames]{xcolor}
\usepackage{hepunits}

\definecolor{bla}{HTML}{03396C}
\definecolor{blaa}{HTML}{005B96}
\definecolor{blaaa}{HTML}{6497B1}

\hypersetup{
  colorlinks, 
  bookmarksopen, 
  bookmarksnumbered,
  citecolor=blaa, 		
  linkcolor=blaa,    	
  urlcolor=blaa,			
}

\usepackage{natbib}

\newcommand{\nf}{\ensuremath{n_f}}

\newcommand{\CA}{\ensuremath{C_A}}
\newcommand{\CF}{\ensuremath{C_F}}
\newcommand{\TF}{\ensuremath{T_F}}

\newcommand{\MS}{\ensuremath{{\rm \overline{MS}}}}
\newcommand{\SMOM}{\ensuremath{{\rm SMOM}}}

\newcommand{\aMS}{\ensuremath{a_{\MS}}}

\newcommand{\tr}{\ensuremath{{\rm tr}}}

\newcommand{\ZpsiR}{\ensuremath{Z_{\psi}^{\rm R}}}
\newcommand{\ZpsiMS}{\ensuremath{Z_{\psi}^{\MS}}}
\newcommand{\ZpsiMOM}{\ensuremath{Z_{\psi}^{\SMOM}}}
\newcommand{\ZpsiMOMmu}{\ensuremath{Z_{\psi}^{\SMOM_{\gamma_\mu}}}}

\newcommand{\ZmR}{\ensuremath{Z_m^{\rm R}}}
\newcommand{\ZmMS}{\ensuremath{Z_m^{\MS}}}
\newcommand{\ZmMOM}{\ensuremath{Z_m^{\SMOM}}}

\newcommand{\mB}{\ensuremath{m_{\rm bare}}}
\newcommand{\mR}{\ensuremath{m_q^{\rm R}}}
\newcommand{\mRI}{\ensuremath{m_q^{\rm RI}}}

\newcommand{\mMS}{\ensuremath{m_q^{\MS}}}
\newcommand{\mMOM}{\ensuremath{m_q^{\SMOM}}}
\newcommand{\CmMStoMOM}{\ensuremath{C_m^{\SMOM}}}

\newcommand{\aB}{\ensuremath{a_{\rm bare}}}

\newcommand{\psiB}{\ensuremath{\psi_{\rm bare}}}
\newcommand{\psiR}{\ensuremath{\psi_{\rm R}}}

\newcommand{\sB}{\ensuremath{(\bar \psi \psi)_{\rm bare}}}
\newcommand{\sR}{\ensuremath{\left[\bar \psi \psi\right]_{\rm R }}}

\begin{document}

\title{
  \Large\color{bla} 
  Quark masses: N3LO bridge from ${\rm RI/SMOM}$ to $\MS$ scheme
}

\author{Alexander {\sc Bednyakov}}\email{bednya@theor.jinr.ru} 
\affiliation{Bogoliubov Laboratory of Theoretical Physics, Joint Institute for Nuclear Research,
Joliot-Curie 6, Dubna 141980, Russia}
\affiliation{P.N. Lebedev Physical Institute of the Russian Academy of Sciences, Leninskii pr., 5, Moscow 119991, Russia}
\author{Andrey {\sc Pikelner}}\email{pikelner@theor.jinr.ru} 
\affiliation{Bogoliubov Laboratory of Theoretical Physics, Joint Institute for Nuclear Research,
Joliot-Curie 6, Dubna 141980, Russia}

\begin{abstract}
	We analytically compute the three-loop corrections to the relation between the renormalized quark masses defined in the minimal-subtraction ($\MS$) and the regularization-invariant symmetric momentum-subtraction (RI/SMOM) schemes.
	Our result is valid in the Landau gauge and 
	can be used 
	to reduce the uncertainty in a lattice determination of the $\MS$ quark masses.

\end{abstract}
\maketitle

\section{Introduction}

Quark masses $m_q$ arise in the Standard Model (SM) from Yukawa interactions of the quarks with the Higgs field. Although not being of fundamental origin, quark masses are usually treated as parameters of the SM and for many years were the only source of information on the Higgs Yukawa couplings. 
As a consequence, precise knowledge of $m_q$ is required both to test the SM and study new physics. The values of the quark masses can be determined 
in several 
ways (for a review see, e.g.,  Ref.~\cite{Tanabashi:2018oca}).   
Since all colored fermions but the top are confined inside hadrons, there is no unique (``physical'') definition of the corresponding mass parameters, and one is free to choose a renormalization scheme that suits better for a problem at hand.
To compare the results of different determinations, it is customary to use perturbation theory (PT) and convert the obtained values to the short-distance running mass $\mMS(\mu)$ 
in the minimal-subtraction scheme $\MS$, evaluated at a fixed scale $\mu$.

\begin{figure}[h]
  \centering
  \begin{tabular}{cc}
    \includegraphics{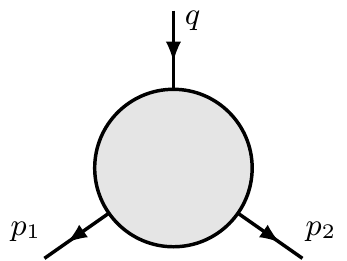} \hspace{2cm} & \includegraphics{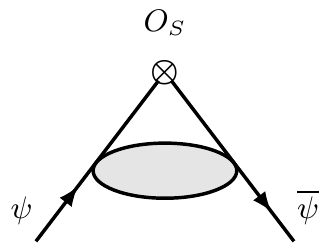}
  \end{tabular}
  \caption{Momentum flow of a Green function (left), and the three-point vertex with $O_S = \bar \psi \psi$ operator insertion (right) considered in the paper. SMOM kinematics corresponds to $p_1^2 = p_2^2 = q^2$, while in the ``exceptional'' case $p_1^2=p_2^2$, and $q^2=0$.}
  \label{fig:v3pt}
\end{figure}

One of the approaches to the quark-mass determination, especially useful in the case of light quarks, is based on lattice computations (see, e.g., Ref.~\cite{Aoki:2019cca}). 
The resulting values, in this case, are bare quark masses $\mB$, corresponding to a particular discretization of QCD with the lattice spacing $a$ acting as the ultraviolet cutoff.
While it is, in principle, possible to directly relate $\mB$ to $\mMS$, it turns out to be more convenient to relate $\mB$ to a mass parameter $\mRI$ defined 
in a regularization-independent (RI) momentum-subtraction renormalization scheme, 
which can be realized directly in lattice QCD. 
The continuum PT is used in this case to convert the finite value $\mRI$ to $\mMS$. 
Among such kind of schemes, the so-called RI/SMOM \cite{Sturm:2009kb}, in which
certain three-point Green functions 
with momenta $p_1$, $p_2$, and $q = p_1 + p_2$ (see, Fig.~\ref{fig:v3pt})
are normalized at \emph{symmetric} kinematics ($p_1^2 = p_2^2 = q^2 = -\mu^2$) 
and have advantages over original RI/MOM
\cite{Martinelli:1994ty} scheme.  
The latter utilizes ``exceptional'' momenta configuration with $q^2=0$, $p_1^2=p_2^2=-\mu^2$ and suffers from enhanced sensitivity to nonperturbative infrared effects (see, e.g., Ref.~\cite{Aoki:2007xm} for details). In addition, the RI/SMOM PT series show a much better convergence behavior than that of the RI/MOM ones. 

Recent state-of-the-art lattice determination \cite{Lytle:2018evc} of the running $\MS$ masses of the charm ($m_c^{\MS}(3~{\rm GeV}) = 0.9896(61)$ GeV) and strange ($m_s^{\MS}(3~{\rm GeV}) = 0.008536(85)$ GeV) quarks in $\nf{}=4$  QCD  
heavily relies on the two-loop (next-to-next-to-leading, or NNLO) conversion factor \cite{Gorbahn:2010bf,Almeida:2010ns} relating $\MS$ and $\SMOM$ schemes. 
According to the estimates given in this reference, the uncertainty due to the missing next-to-next-to-next-to-leading (N3LO) term is comparable with other sources of uncertainties (e.g., due to continuum extrapolation or condensate effects) and contribute a significant part to the overall error budget (for details 
see Table VI of Ref.\cite{Lytle:2018evc}). 

In this letter,  we report on the analytical computation of the three-loop contribution, thus, providing additional 
precision for such an analysis. Recently, a \emph{numerical} evaluation of the same quantity appeared in Ref.~\cite{Kniehl:2020sgo}. Our result confirms the estimates provided therein.

\section{Details of calculation}

To calculate the required conversion factor $\CmMStoMOM$, we consider QCD with $\nf{}$ flavors and define 
\begin{align}
	\mMS = \CmMStoMOM \mMOM, 
	\quad 
	\CmMStoMOM = \frac{\ZmMOM}{\ZmMS}. 
	\label{eq:CM_def}
\end{align}
	The mass parameters in $\MS$ and $\SMOM$ schemes are related to the quark bare mass $\mB$ 
via $\ZmR = \{ \ZmMS, \ZmMOM \}$
\begin{align}
	\mB = \ZmR \mR = \ZmMS \mMS = \ZmMOM \mMOM.
	\label{eq:m_bare}
\end{align}
In continuum QCD the bare mass $\mB$ is usually defined in dimensional regularization so that each $\ZmR$ contains poles in $\varepsilon = (4-d)/2$.   To determine 
$\ZmR$ we do not compute massive propagators but renormalize the scalar bilinear operator $O_S \equiv \bar \psi \psi$ (see Fig.~\ref{fig:v3pt}) in massless QCD  
\begin{align}
	\sR = \ZmR \sB. 
	\label{eq:Sop_R}
\end{align}
This simplified approach neglects both valence and sea quark masses, but still provides a reasonable approximation to the conversion factor $\CmMStoMOM$ in a range of renormalization scales utilized in lattice calculations (see,e.g,  Ref.~\cite{Lytle:2018evc} for numerical studies of the two-loop corrections due to nonzero quark masses). 
	
We compute $\ZmMOM$ and $\ZmMS$ order-by-order in PT by considering
	bare three-point one-particle-irreducible vertex function 
	\begin{align}
		\left.\Lambda_S(p_1,p_2)\right|_{sym} =  
			\left.\langle \psi(-p_2) O_S(q) \bar \psi(-p_1) \rangle\right|_{ p_1^2 = p_2^2  = q^2  = -\mu^2}, \quad q = p_1 + p_2
				\label{eq:GF_def}
	\end{align}
	in $\SMOM$ kinematics. We use Landau gauge and require that 
\begin{align}
	1  = \ZmMOM \cdot \ZpsiMOM \cdot  \frac{1}{12} \cdot  \left. \tr \left[ \Lambda^{\rm bare}_S \right]\right|_{sym}, 
		\quad
	1  = \ZpsiMOM \cdot \frac{1}{12 p^2} \left. \cdot \tr \left[ i S_{\rm bare}^{-1}(p) \hat p \right]\right|_{p^2 = - \mu^2}, 
		\label{eq:SMOM_RC}
	\end{align}
where both $\Lambda_S^{\rm bare}$ and the bare quark inverse propagator 
$S_{\rm bare}^{-1}$ 
are reexpanded in terms of $\MS$ strong coupling $\alpha_s^{\MS} = (4\pi) \aMS$ via the well-known formula $\mu^{-2 \varepsilon} \aB = Z_{\aMS} \aMS$ available with five-loop accuracy \cite{Chetyrkin:2017bjc,Luthe:2017ttg}. 
In Eq.~\eqref{eq:SMOM_RC} the quark field renormalization constants are defined as\footnote{It is worth mentioning that, e.g., in Refs.~\cite{Sturm:2009kb,Almeida:2010ns,Lytle:2018evc}, different notation can be adopted for the renormalization constants,and one should make the substitutions  $Z_\psi \to Z_\psi^{-1}$ and $Z_m \to Z^{-1}_m$ to compare the results.} 
\begin{align}
	\psiB = \sqrt{\ZpsiR} \psiR, \quad {\rm R} = \{ \MS, \SMOM \}  
	\label{eq:psi_bare}
\end{align}

The conditions \eqref{eq:SMOM_RC} can be implemented in lattice computations, leading to a nonperturbative determination \cite{Martinelli:1994ty} of $\ZmMOM$. The latter converts the bare lattice mass into $\mMOM$, providing input for $\mMS$ calculation via Eq.~\eqref{eq:CM_def}. 
The $\MS$ counterparts $\ZmMS$, $\ZpsiMS$ of the renormalization constants 
in Eq.~\eqref{eq:SMOM_RC} required to compute $\CmMStoMOM$ are obtained by subtracting only divergent terms of the corresponding Green functions.

A comment is in order regarding the determination of the wave function renormalization constant $Z_\psi^{\SMOM}$. Due to Ward identities, the latter can also be obtained from the (non)renormalization of vector (axial) quark bilinear operators $O^\mu_V \equiv \psi \gamma^\mu \psi$ ($O^\mu_A \equiv \psi \gamma^\mu \gamma_5\psi$). 
In the continuum, Ward identifies and chiral symmetry guarantee that $Z_V = Z_A = 1$, and it can be proven \cite{Sturm:2009kb} that the condition on $Z_\psi^{\SMOM}$ given in Eq.~\eqref{eq:SMOM_RC}  corresponds to 
\begin{align}
	1  = \ZpsiMOM \cdot  \frac{1}{12 q^2} \cdot  \left. \tr \left[ q_\mu \Lambda^{\mu,\rm bare}_{V} \hat q  \right]\right|_{sym}, \quad
	1  = \ZpsiMOM \cdot  \frac{1}{12 q^2} \cdot  \left. \tr \left[ q_\mu \Lambda^{\mu,\rm bare}_{A} \gamma_5 \hat q  \right]\right|_{sym} 
		\label{eq:SMOM_wf_RC}
\end{align}
with $\Lambda_V^\mu$ ($\Lambda_A^\mu$) being analogs of \eqref{eq:GF_def} with $O_S$ replaced by $O_V^\mu$ ($O_A^\mu$). 
It is also possible to use the so-called RI/SMOM${}_{\gamma_\mu}$ \cite{Sturm:2009kb} and require
\begin{align}
	1  = \ZpsiMOMmu \cdot  \frac{1}{48} \cdot  \left. \tr \left[ \gamma_\mu \Lambda^{\mu,\rm bare}_{V}  \right]\right|_{sym}, \quad
	1  = \ZpsiMOMmu \cdot  \frac{1}{48} \cdot  \left. \tr \left[ \Lambda^{\mu,\rm bare}_{A} \gamma_5 \gamma_\mu  \right]\right|_{sym}.
		\label{eq:SMOMmu_wf_RC}
\end{align}
Both RI/SMOM and RI/SMOM${}_{\gamma_\mu}$ conditions can be implemented on lattice (see, e.g., Refs.~\cite{Blum:2014tka,Aoki:2007xm} for details and subtleties). In Ref.~\cite{Almeida:2010ns} it was demonstrated that the PT series for the quark-mass conversion factor exhibits slightly better behavior in RI/SMOM than in RI/SMOM${}_{\gamma_\mu}$. Given this argument we carry out our calculation in RI/SMOM.

Let us mention a few technical details of our calculation.  We generate Feynman graphs with \texttt{DIANA}~\cite{Tentyukov:1999is} 
and take fermion and color \cite{vanRitbergen:1998pn} traces according to
Eq.~\eqref{eq:SMOM_RC}. Resulting scalar integrals are reduced to the set of master
integrals identified in our previous paper~\cite{Bednyakov:2020cdf} on $\alpha_s$
renormalization in the $\SMOM$ scheme. To perform reduction we make use of the
\texttt{FIRE6}\cite{Smirnov:2019qkx} package. Substituting masters integrals
evaluated previously, we end up with expressions valid for a general gauge group.
The number of master integrals and the necessary expansion depth in dimensional
regularization parameter $\varepsilon=(4-d)/2$ are the same as in the paper\cite{Bednyakov:2020cdf}. 
It is worth noting that as a cross-check of our calculation we also consider the renormalization of the  pseudoscalar quark current $O_P=\bar \psi \gamma_5 \psi$, which can also be used to extract $\ZmMOM$ from lattice calculations.  

\section{Results and conclusion}

Expressing all the renormalization constants in terms of $\aMS$, from 
Eq.~\eqref{eq:CM_def} we obtain the following N3LO conversion factor
\begin{align}
	\CmMStoMOM & = 1 + x_1 \aMS  + x_2 \aMS^2 + x_3 \aMS^2 
	\label{eq:cm_res}
\end{align}
with
{\allowdisplaybreaks
  \begin{align}
    x_1 = &   \textcolor{bla}{\CF}
            \bigg(
            -4
            - \frac{2}{3} \pi^2
            + \psi_1
            \bigg) \label{eq:cm_1}\\
    x_2 = & \textcolor{bla}{\nf \TF \CF}
            \bigg(
            \frac{83}{6}
            + \frac{40}{27} \pi^2
            - \frac{20}{9} \psi_1
            \bigg)
            \nonumber\\
    + & \textcolor{bla}{\CF^{2}}
        \bigg(
        \frac{19}{8}
        + \frac{28}{9} \pi^2
        - \frac{14}{3} \psi_1
        + 4 \zeta_3
        + \frac{58}{81} \pi^4
        - \frac{52}{27} \psi_1 \pi^2
        + \frac{13}{9} \psi_1^2
        - \frac{1}{36} \psi_3
        \bigg)
        \nonumber\\
    + & \textcolor{bla}{\CA \CF}
        \bigg(
        - \frac{1285}{24}
        - \frac{385}{54} \pi^2
        + \frac{385}{36} \psi_1
        + 10 \zeta_3
        - \frac{8}{81} \pi^4
        + \frac{8}{27} \pi^2 \psi_1 
        - \frac{2}{9} \psi_1^2
        \bigg)\label{eq:cm_2}\\
    x_3 = & \textcolor{bla}{\nf^{2} \TF^{2} \CF}
            \bigg(
            -\frac{7514}{243}
            - \frac{800}{243} \pi^2
            + \frac{400}{81} \psi_1
            - \frac{32}{9} \zeta_3
            - \frac{32}{243} \pi^4
            + \frac{4}{81} \psi_3
            \bigg)
            \nonumber\\
    + & \textcolor{bla}{\nf \TF \CA \CF}
        \bigg(
        \frac{95387}{243}
        + \frac{13172}{243} \pi^2
        - \frac{6586}{81} \psi_1
        - \frac{152}{9} \zeta_3
        + \frac{3952}{3645} \pi^4
        \nonumber\\
    - & \frac{320}{243} \psi_1 \pi^2
        + \frac{80}{81} \psi_1^2
        - \frac{23}{162} \psi_3
        + \frac{320}{81} \pi^2 \zeta_3
        + \frac{16240}{729} \zeta_5
        - \frac{160}{27} \psi_1 \zeta_3
        + \frac{64}{81} H_5
        \bigg)
        \nonumber\\
    + & \textcolor{bla}{\nf \TF \CF^2}
        \bigg(
        \frac{1109}{9}
        - \frac{241}{81} \pi^2
        + \frac{241}{54} \psi_1
        - \frac{1384}{9} \zeta_3
        - \frac{15392}{3645} \pi^4
        + \frac{2080}{243} \psi_1 \pi^2
        \nonumber\\
    - & \frac{520}{81} \psi_1^2
        + \frac{67}{162} \psi_3
        - \frac{128}{9} \pi^2 \zeta_3
        - \frac{32480}{729} \zeta_5
        + \frac{64}{3} \psi_1 \zeta_3
        - \frac{128}{81} H_5
        \bigg)
        \nonumber\\
    + & \textcolor{bla}{\CF^{3}}
        \bigg(
        - \frac{3227}{12}
        - \frac{191}{12} \pi^2
        + \frac{191}{8} \psi_1
        -58 \zeta_3
        - \frac{992}{81} \pi^4
        + \frac{232}{9} \psi_1 \pi^2
        - \frac{58}{3} \psi_1^2
        + \frac{37}{27} \psi_3
        \nonumber\\
    + & \frac{80}{9} \pi^2 \zeta_3
        - \frac{32980}{81} \zeta_5
        - \frac{40}{3} \psi_1 \zeta_3
        - \frac{112}{9} H_5
        - \frac{131776}{98415} \pi^6
        + \frac{23992}{6561} \psi_1 \pi^4
        - \frac{394}{81} \psi_1^2 \pi^2
        \nonumber\\
    + & \frac{679}{6561} \psi_3 \pi^2
        - \frac{679}{4374} \psi_1 \psi_3
        + \frac{197}{81} \psi_1^3
        + \frac{1}{135} \psi_5
        + \frac{2}{8505} H_6
        \bigg)
        \nonumber\\
    + & \textcolor{bla}{\CA \CF^{2}}
        \bigg(
        \frac{18781}{72}
        + \frac{23231}{324} \pi^2
        - \frac{23231}{216} \psi_1
        + \frac{2879}{9} \zeta_3
        + \frac{34423}{1458} \pi^4
        - \frac{11306}{243} \psi_1 \pi^2
        \nonumber\\
    + & \frac{5653}{162} \psi_1^2
        - \frac{3937}{1296} \psi_3
        - \frac{178}{9} \pi^2 \zeta_3
        + \frac{379285}{729} \zeta_5
        + \frac{89}{3} \psi_1 \zeta_3
        + \frac{1840}{81} H_5
        - \frac{1519}{32805} \pi^6
        \nonumber\\
    + & \frac{4}{81} \psi_1 \pi^4
        + \frac{4}{27} \psi_1^2 \pi^2
        + \frac{1}{27} \psi_3 \pi^2      
        - \frac{1}{18} \psi_1 \psi_3
        -   \frac{2}{27} \psi_1^3
        - \frac{77}{116640} \psi_5      
        \bigg)
        \nonumber\\
    + & \textcolor{bla}{\CA^{2} \CF}
        \bigg(
        -\frac{3360023}{3888}
        - \frac{243283}{1944} \pi^2
        + \frac{243283}{1296} \psi_1
        + \frac{4511}{24} \zeta_3
        - \frac{20513}{5832} \pi^4
        + \frac{5107}{972} \psi_1 \pi^2
        \nonumber\\
    - & \frac{5107}{1296} \psi_1^2
        + \frac{3433}{5184} \psi_3
        + \frac{7535}{324} \pi^2 \zeta_3
        - \frac{1140715}{5832} \zeta_5
        - \frac{7535}{216} \psi_1 \zeta_3
        - \frac{668}{81} H_5          
        + \frac{100133}{314928} \pi^6\nonumber\\
    - & \frac{10619}{13122} \psi_1 \pi^4
        + \frac{325}{324} \psi_1^2 \pi^2
        - \frac{461}{13122} \psi_3 \pi^2
        +  \frac{461}{8748} \psi_1 \psi_3          
        - \frac{325}{648} \psi_1^3
        - \frac{611}{373248} \psi_5
        - \frac{1}{17010} H_6
        \bigg).
        \label{eq:cm_3}
  \end{align}
}
Here $\zeta_i$ is the Riemann zeta function, and $\psi_m = \psi^{(m)}(1/3)$
corresponds to the $(m+1)$th
derivative of the gamma function. Additional constants of uniform transcendental
weight $H_5$ and $H_6$, introduced in Ref.~\cite{Bednyakov:2020cdf},
\begin{equation}
  \label{eq:H5H6num}
  H_5 = - 23.9316195698,   \quad H_6 =   248215.038289
\end{equation}
are linear combinations of real parts of harmonic polylogarithms with six-root of unity
argument from the basis constructed in Ref.~\cite{Kniehl:2017ikj}.
Our result reproduces the well-known analytic one-loop \cite{Sturm:2009kb} and two-loop \cite{Gorbahn:2010bf,Almeida:2010ns} expressions, together with recent numerical evaluation of Ref.~\cite{Kniehl:2020sgo}:

\begin{align}
  \label{eq:CmNF}
  \CmMStoMOM = 1 & - 0.6455188560 \aMS  -  (22.60768757 - 4.013539470 \nf)\aMS^2 \nonumber\\
                 & - (860.2874030 - 164.7423004 \nf + 2.184402262 \nf^2) \aMS^3.
\end{align}
%

Given this general result (\ref{eq:CmNF}), we are ready to provide our numerical
estimates of the N3LO contribution for different $\nf$. 
Expanding the matching factor in powers of $\alpha_s \equiv \alpha_s^{\MS}$, we obtain
\begin{alignat}{3}
  \label{eq:nf4alpha}
  \nf =0 :\quad& 1 &&- 0.05136875839 \alpha_s -0.1431648540 \alpha_s^2 &&- 0.4335248250 \alpha_s^3,
  \\
  \nf=1 :\quad& 1 &&- 0.05136875839 \alpha_s - 0.1177488184 \alpha_s^2 && - 0.3516069867 \alpha_s^3,
  \\
  \nf=2 :\quad& 1 &&- 0.05136875839 \alpha_s - 0.09233278278 \alpha_s^2 && - 0.2718907211 \alpha_s^3,
  \\
  \nf=3 :\quad& 1 &&- 0.05136875839 \alpha_s - 0.06691674717 \alpha_s^2 && - 0.1943760281 \alpha_s^3,
  \\
  \nf=4 :\quad& 1 &&- 0.05136875839 \alpha_s - 0.04150071157 \alpha_s^2 && - 0.1190629077 \alpha_s^3,
  \\
  \nf=5 :\quad& 1 &&- 0.05136875839 \alpha_s - 0.01608467597 \alpha_s^2 && - 0.04595136006 \alpha_s^3,
  \\
  \nf=6 :\quad& 1 &&- 0.05136875839 \alpha_s +0.009331359638 \alpha_s^2 && + 0.02495861498 \alpha_s^3.
\end{alignat}

Given the value $\alpha_s^{nf=4}(3\,\GeV) = 0.2545$ used by HPQCD collaboration~\cite{Lytle:2018evc} in
the determination of charm- and strange-quark masses, we evaluate the matching factor 
at the reference scale $\mu_{\rm ref} = 3\,\GeV$
\begin{align}
  \label{eq:nf4alpha}
  Z^{\MS/\SMOM}_m \equiv \CmMStoMOM & = 1 
  - \underbrace{0.0130733}_{\alpha_s} 
  - \underbrace{0.00268801}_{\alpha_s^2} 
  - \underbrace{0.00196264}_{\alpha_s^3} = 0.982276,
  \quad
\nf=4,~\mu=3~\GeV.
\end{align}
One can see that 
the three-loop contribution is of the same order as the two-loop correction and is of the same size as the uncertainty $0.22\%$ 
quoted in Ref.~\cite{Lytle:2018evc} and attributed to the missing N3LO term. 
The comparision with the result given in Ref.~\cite{Lytle:2018evc} also shows that
the effect of the $\alpha_s^3$ term in Eq.~\eqref{eq:nf4alpha} is four times larger than the two-loop contribution due to massive charm quark in the sea and becomes an order of magnitude larger if $\mu=5$ GeV is chosen.

It is also worth mentioning that the authors of Ref.~\cite{Kniehl:2020sgo} also
consider vector and tensor quark bilinears. We apply the projector
\eqref{eq:SMOMmu_wf_RC} to the expression for the vector-operator $O_V$ matrix
element given in Ref.~\cite{Kniehl:2020sgo}, evaluate the quark wave function renormalization in RI/SMOM${}_{\gamma_\mu}$, and obtain the following numeric result for the corresponding matching factor: 
\begin{align}
	C_m^{\SMOM_{\gamma_\mu}}  = 1  - & 1.978852189 \aMS - (  55.03243483 - 6.161687618 \nf) \aMS^2 \nonumber\\
	- & (2086.34(14) - 362.560(3) \nf + 6.7220(1) \nf^2) \aMS^3.
	\label{eq:CmMuNF}
\end{align}
While the two-loop contribution to Eq.~\eqref{eq:CmMuNF} is known in analytic form \cite{Almeida:2010ns}, the three-loop term is new and, to our knowledge, is not presented in the literature.  One can see that numerical coefficients in RI/SMOM${}_{\gamma_\mu}$ \eqref{eq:CmMuNF} is indeed larger than that in RI/SMOM~\eqref{eq:CmNF}, and, e.g., at our reference scale $\mu_{\rm ref}$ we have 
\begin{align}
  \label{eq:nf4alphaMu}
 C_m^{\SMOM_{\gamma_\mu}} & = 1 
  - \underbrace{0.04007663}_{\alpha_s} 
  - \underbrace{0.012463065}_{\alpha_s^2} 
  - \underbrace{0.006177}_{\alpha_s^3} = 0.941283,
  \quad
\nf=4,~\mu=3~\GeV.
\end{align}

To conclude, we analytically calculate the three-loop correction to the matching factor in RI/SMOM scheme required to extract $\MS$ quark masses from nonperturbative lattice computations\cite{Blum:2014tka,Lytle:2018evc}. Our numerical evaluation confirms the estimate of $x_3$ given in Ref.~\cite{Kniehl:2020sgo}. In addition, we use the results of Ref.~\cite{Kniehl:2020sgo} to evaluate the three-loop expression for the corresponding matching factor in RI/SMOM${}_{\gamma_\mu}$. We believe that the obtained N3LO contribution to $\CmMStoMOM$ will increase the precision of the resulting $\MS$ quark masses and/or provide a more reliable estimate of the uncertainties due to missing high-order terms.

\acknowledgments

We would like to thank Christine Davies for the correspondence regarding Ref.~\cite{Lytle:2018evc} and clarifying comments on the sea quark contribution.
The work of A.P. is supported by the Foundation for the Advancement
of Theoretical Physics and Mathematics ``BASIS.''
The work of A.B. is supported by the Grant of the Russian Federation Government, Agreement No. 14.W03.31.0026 from 15.02.2018.
\bibliography{mqSMOM}
\end{document}